\documentstyle[epsfig,12pt]{article}
\newcommand{\gsim}{\raisebox{-0.07cm}{$\, \stackrel{>}{{\scriptstyle
\sim}}\, $}}

\begin{document}
\thispagestyle{empty}

\phantom{X}
\vspace{2.5cm}

\begin{center}
{\Large\bf Vector Meson Production at a Polarized HERA Collider}

\vspace{1cm}
{\large S.P.~Baranov$^a$, M.G.~Ryskin$^b$, T.~Gehrmann$^c$}

\vspace*{1cm}
{\it $^a$P.N.~Lebedev Physical Institute, Moscow 117924, Russia}\\

\vspace*{3mm}
{\it $^b$Petersburg Nuclear Physics Institute,
188350, Gatchina, St.Petersburg, Russia}\\

\vspace*{3mm}
{\it $^c$DESY Theory Group, DESY, D-22603 Hamburg, Germany}\\

\vspace*{2cm}

\end{center}

\begin{center}
{\bf Abstract}
\end{center}
A successful interpretation of the cross sections for elastic and
inelastic vector meson production can be made if one assumes the
underlying dynamics to be governed by hard partonic subprocesses.
Extending these partonic approaches to longitudinally polarized
electron-proton collisions, we estimate the expected
production asymmetries at the HERA collider. The anticipated
statistical errors on these asymmetries mostly turn out to be larger
than the asymmetries themselves, such that
an experimental
observation of these asymmetries at HERA looks not feasible.

\vspace{6.5cm}
\noindent
{\it Contribution to the Proceedings of the ``Workshop on 
Physics with Polarized Protons at HERA'', DESY, March--September 1997.}
\vfill

\setcounter{page}{0} 
\newpage

\begin{center}
{\Large\bf Vector Meson Production at a Polarized HERA Collider}

\vspace{1cm}
{\large S.P.~Baranov$^a$, M.G.~Ryskin$^b$, T.~Gehrmann$^c$}

\vspace*{1cm}
{\it $^a$P.N.~Lebedev Physical Institute, Moscow 117924, Russia}\\

\vspace*{3mm}
{\it $^b$Petersburg Nuclear Physics Institute,
188350, Gatchina, St.Petersburg, Russia}\\

\vspace*{3mm}
{\it $^c$DESY Theory Group, DESY, D-22603 Hamburg, Germany}\\

\vspace*{2cm}

\end{center}

\begin{abstract}
A successful interpretation of the cross sections for elastic and
inelastic vector meson production can be made if one assumes the
underlying dynamics to be governed by hard partonic subprocesses.
Extending these partonic approaches to longitudinally polarized
electron-proton collisions, we estimate the expected
production asymmetries at the HERA collider. The anticipated
statistical errors on these asymmetries mostly turn out to be larger
than the asymmetries themselves, such that
an experimental
observation of these asymmetries at HERA looks not feasible.
\end{abstract}

\section{Introduction}
\label{sec:intro}
\vspace{1mm}
\noindent

Vector mesons produced in lepton-nucleon collisions can be
identified relatively easy
from their decay signatures (e.g.~$\rho \to \pi^+\pi^-$
or $J/\psi \to \mu^+\mu^-$). They are therefore among the most commonly
studied final state observables in lepton-nucleon scattering
experiments (e.g.~\cite{h1exp,zeusexp}).
Depending on the virtuality of the incoming photon, one distinguishes
the photoproduction ($Q^2\approx 0$) and leptoproduction ($Q^2 \gg
\Lambda^2_{{\rm QCD}}$)
of vector mesons. The photoproduction cross sections are usually
larger than the leptoproduction cross sections, which
show a steep fall--off with increasing $Q^2$.
Moreover, one has to distinguish between elastic (diffractive)
and inelastic production of vector mesons. The identification of the
elastic events in the experimental measurement can be made by
tagging on a large rapidity gap around the outgoing proton direction or
by requiring the vector meson to have the same energy as the incoming photon.

Both elastic and inelastic cross sections can be modeled in approaches
based on a partonic interpretation of the underlying subprocess. The
elastic production of vector mesons can be interpreted to be due to
perturbative two gluon exchange with a $q\bar q$ pair in the
photon wave function, which subsequently forms the
vector meson~\cite{BLFS,R1}. This model and its predictions for
elastic polarization
asymmetries will be presented in Section~\ref{sec:elth} and numerical
predictions for HERA are contained in Section~\ref{sec:elnum}. The color
singlet model~\cite{colsing} for
inelastic heavy vector meson production assumes this process
to be dominated by  photon-gluon fusion, yielding a $q\bar q$ final state
which transforms into a color singlet state by the emission of
 one hard gluon. Implications of this model for polarization asymmetries
will be discussed in Section~\ref{sec:inelth}.
For both elastic and inelastic production, we estimate the expected
statistical errors of a measurement at a polarized HERA collider
with an integrated luminosity of 1000~pb$^{-1}$, based on the measured
unpolarized cross sections.

\section{Two gluon exchange model for elastic vector meson production}
\label{sec:elth}
\vspace{1mm}
\noindent
The final state configuration of diffractive electron-proton interactions,
containing an only minimally deflected proton, requires these processes
to be mediated by the exchange of a color singlet configuration between
proton and virtual photon.
The simplest purely  perturbative model for this color singlet exchange
is given by two gluons, which are emitted simultaneously from the
incoming proton. This perturbative two gluon exchange is the basis of
a model~\cite{BLFS,R1}  which yields a successful description of the
unpolarized elastic $J/\psi$ and $\rho$ production cross sections.

The leading logarithmic contribution to diffractive vector meson
production in this model
comes from the Feynman graphs shown in
Fig.1. The upper quark loop describes the transition of 
the initial photon into
 a quark--antiquark pair which then forms the vector meson $V$.
The vector meson wave function is
included into the vertex $q\bar{q}\to V$.
\begin{figure}[b!] % fig 1
\begin{center}
~ \epsfig{file=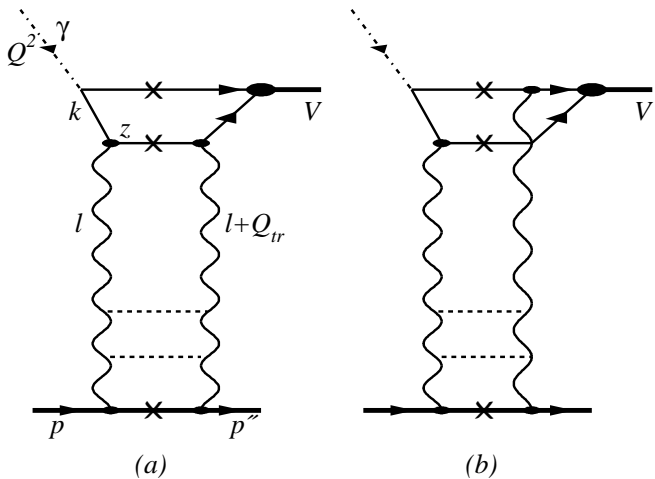,width=9cm}
\caption{
 Diffractive open $q\bar{q}$ production in high energy $\gamma^*p$
collisions via the two gluon exchange.}
\label{fig1}
\end{center}
\end{figure}

In the Born approximation (neglecting the dashed lines in Fig.1) the
amplitude of the interaction of the $q\bar{q}$ pair with the target proton is
given by the two gluon exchange. We consider the region of small
$x=(Q^2+M^2_V)/W^2 \ll 1$ and large scale $Q^2+M^2_V\gg \Lambda^2$.
The last is needed to justify the perturbative QCD approach.

To find  the  leading logarithmic approximation (LLA),
which resums all terms  of the form
$\alpha_s^n\cdot \ln^n[(Q^2+M^2_V)/\Lambda^2]$,
one has to consider the ladder type diagrams which include the dashed
 lines shown in Fig.1. This is done by the
replacement of the lower part of the diagram by the gluon structure function
$G(x,q^2)$ which describes the probability to find an appropriate $t$-channel
gluon in a proton. Therefore one finds for the unpolarized
amplitude ${\cal A}\propto G(x,q^2)$. Extending this calculation to the
spin dependent part of the amplitude, one finds 
$\Delta {\cal A}\propto x\Delta G$~\cite{R2,R3}.

From the formal point of view, 
the amplitudes ${\cal A}$ and $\Delta {\cal A}$ are
proportional to the off-diagonal distributions $G(x,x')$ (or $\Delta G(x,x')$)
but when the difference $\delta x=x-x'\ll 1$  is small these off-diagonal
distribution are rather close \cite{FS}
 ($\pm 10\ldots20$\%) to the diagonal ones, i.e.~to
the conventional structure functions $G(x)$ and $\Delta G(x)$.

The asymmetry for the $J/\psi$ photo(lepto)production was calculated in
\cite{R2} in the leading order (LO) LLA using the non-relativistic $J/\psi$
wave function. It was found to be
\begin{equation}
A^{\gamma^* p \to J/\psi p}
=\frac{\sigma^{\uparrow\downarrow}-\sigma^{\uparrow\uparrow}}
{\sigma^{\uparrow\downarrow}+\sigma^{\uparrow\uparrow}}\;=\;
\frac{2\Delta G\cdot G}{G^2+\Delta G^2}\simeq\frac{2\Delta G}{G}\; ,
\end{equation}
where the factor 2 arises from
 the fact that the unpolarized cross section is proportional to
the gluon density squared while the polarized cross section comes from
the interference term of spin-dependent and spin-independent amplitudes.

While the large charm quark mass justified the use of a
non-relativistic wave function for the $J/\psi$, this is
clearly inappropriate to $\rho$
production. A consistent calculation would require the incorporation of
the (unknown) relativistic wave function of the
$\rho$-meson. To overcome this problem the hadron-parton duality hypothesis
 was used. The idea is as follows.
Let us consider the
amplitude of the open $q\bar{q}$ leptoproduction, project it on the
state with the spin $J^{PC}=1^{--}$ (isospin $I^G=1^+$)
and average over a mass
interval $\Delta M_{q\bar{q}}$ (typically $\sim 1$ GeV$^2$) in the region of
$\rho$ mass. In this domain the more complicated partonic states
($q\bar{q}+g,\; q\bar{q}+2g,\; q\bar{q}+q\bar{q},...$)
are suppressed, while on
the hadronic side the $2\pi$ states
 are known to dominate. Thus for low $M^2$ we
mainly have $\gamma^*\to q\bar{q}\to 2\pi\to\rho$ or in other words
\begin{equation}
\sigma(\gamma^*p\to\rho p)\simeq \sum_{q=u,d}\int^{M^2_b}_{M^2_a}
\frac{\sigma(\gamma^*p\to (q\bar{q})' p)}{dM^2}dM^2
\end{equation}
where the limits $M^2_a$ and $M^2_b$ are chosen so that they appropriately
 embrace the $\rho$-meson mass region and $(q\bar{q})'$ means the projection of
 the $q\bar{q}$ system onto the $J^{PC}=1^{--}$, $I^G=1^+$ state.
\begin{figure}[b!] % fig 2
\begin{center}
~ \epsfig{file=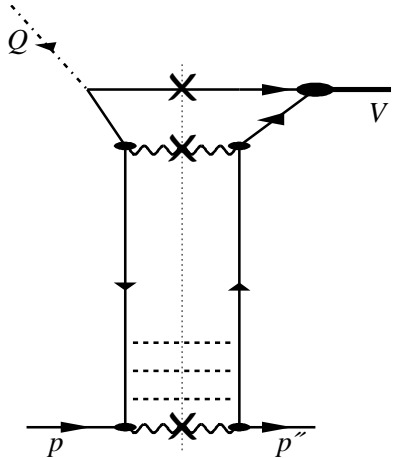,width=5cm}
\caption{
Diffractive open $q\bar{q}$ production in high energy $\gamma^*p$
collisions via the quark exchange.}
\label{fig2}
\end{center}
\end{figure}

In ref.~\cite{MRT} it was demonstrated that  based on this hadron-parton
 duality one can describe the diffractive $\rho$-meson
 electroproduction on the unpolarized target rather well
in the framework of perturbative QCD, for both
longitudinal and transverse $\rho$ polarization.
Thus we use the hadron-parton duality for the polarized cross sections as well.

It turns out that in the relativistic $\rho$ system
  the asymmetry is washed out by the
``Fermi motion'' of the light quarks in the $\rho$-meson wave function. So the
analyzing power becomes 4 times smaller~\cite{R3}. Including the
projection on the $\rho$ quantum numbers, the asymmetry reads
\begin{equation}
A^{\gamma^* p \to \rho p} \simeq \frac{\gamma(G)\Delta G}{2\gamma(\Delta G) G}
\; ,
\end{equation}
where $\gamma(G,\Delta G)$ are the anomalous dimensions of the
unpolarized and polarized gluon distributions in the small $x$ region.
Besides this the spin dependent light quark distributions also contribute to
the $\rho$ asymmetry (see Fig.2).

This approach can be extended a variety of other diffractive processes,
including e.g.~vector meson production with diffractive target dissociation.
A brief overview is given in~\cite{R5}.

\section{Asymmetries in elastic
$\rho$ leptoproduction and $J/\psi$ photoproduction at HERA}
\label{sec:elnum}
\vspace{1mm}
\noindent
Using the formulae for asymmetries in elastic vector meson production
discussed above, it is possible to produce quantitative estimates
for these asymmetries at HERA. Given that the polarized
gluon distribution $\Delta G(x,Q^2)$ is only loosely constrained
from present data, these estimates can only give an idea of the
expected magnitude of the asymmetries, without attempting to make
precise predictions. To illustrate this uncertainty, we use
different parameterizations of the (leading order)
polarized gluon distribution,
GS(A--C)~\cite{GS} and GRSVs~\cite{GRSV}, which were obtained
from fits to polarized deep inelastic scattering data. These are
consistently used in combination with the unpolarized (leading order)
gluon distribution of~\cite{GRV}.
\begin{figure}[b!] % fig 3
\begin{center}
~ \epsfig{file=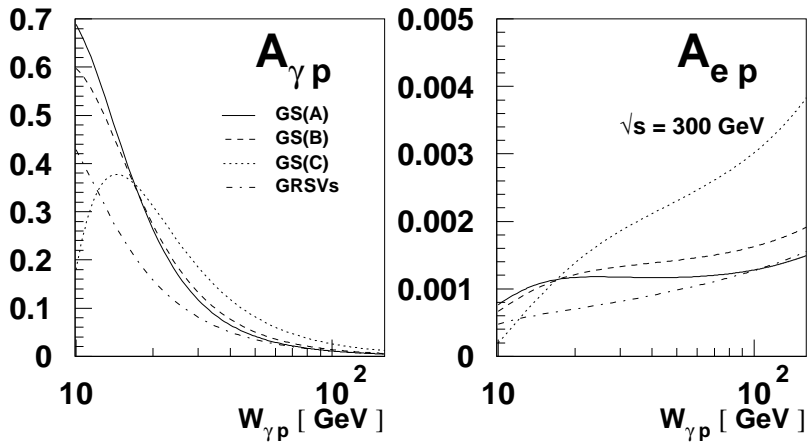,height=11cm,angle=-90}
\caption{Asymmetries in elastic $J/\psi$ photoproduction: photon-proton
asymmetry (left) and electron-proton asymmetry at HERA (right).}
\label{fig3}
\end{center}
\end{figure}

All asymmetries defined above are valid for the ideal case of the
collision of a polarized photon with a polarized proton. In electron-proton
scattering experiments, only part of the electron polarization is transferred
to the photon. This photon depolarization is described by the ratio of
polarized to unpolarized electron-to-photon splitting functions
%\begin{displaymath}
\begin{equation}
D(y) = \frac{1-(1-y)^2}{1+(1-y)^2}\; \label{D}.
\end{equation}
%\end{displaymath}
The effect of the photon depolarization is illustrated in Fig.~\ref{fig3},
comparing the photon-proton and electron-proton asymmetries for
$J/\psi$ photoproduction at HERA. In particular at low values of
$W_{\gamma p}$,
corresponding to low $y$, the asymmetry is lowered by several orders
of magnitude. The resulting observable $J/\psi$ photoproduction
asymmetry in electron-proton collisions never exceeds 0.004. Based on the
experimental measurement of the unpolarized elastic $J/\psi$ photoproduction
cross section~\cite{zeusexp}, it is possible to estimate
the expected statistical accuracy of such a measurement. For an
integrated luminosity of 1000~pb$^{-1}$, product of
beam polarizations $P_p\cdot P_e=0.5$ and reconstruction efficiency
$\epsilon_{J/\psi}=0.12$, we find that it would be possible to obtain four
data-points in the region of  $W_{\gamma p}=40\ldots 140$~GeV,
each with a statistical error of $\pm 0.005$ on the asymmetry.
This estimate illustrates that the elastic $J/\psi$ photoproduction asymmetry
is unlikely to be observable at the HERA collider. We have moreover
checked that no significant improvement is made for this measurement
if the HERA
centre-of-mass energy is lowered to
$\sqrt s=180$~GeV.

Finally,
the electron-proton asymmetry for elastic $\Upsilon$ photoproduction is
found to be about
one order of magnitude larger than for the $J/\psi$. Since the
unpolarized elastic $\Upsilon$ photoproduction cross section is not accurately
known at present, it is however difficult to provide a sensible
error estimate for this asymmetry.
\begin{figure}[t!] % fig 4
\begin{center}
~ \epsfig{file=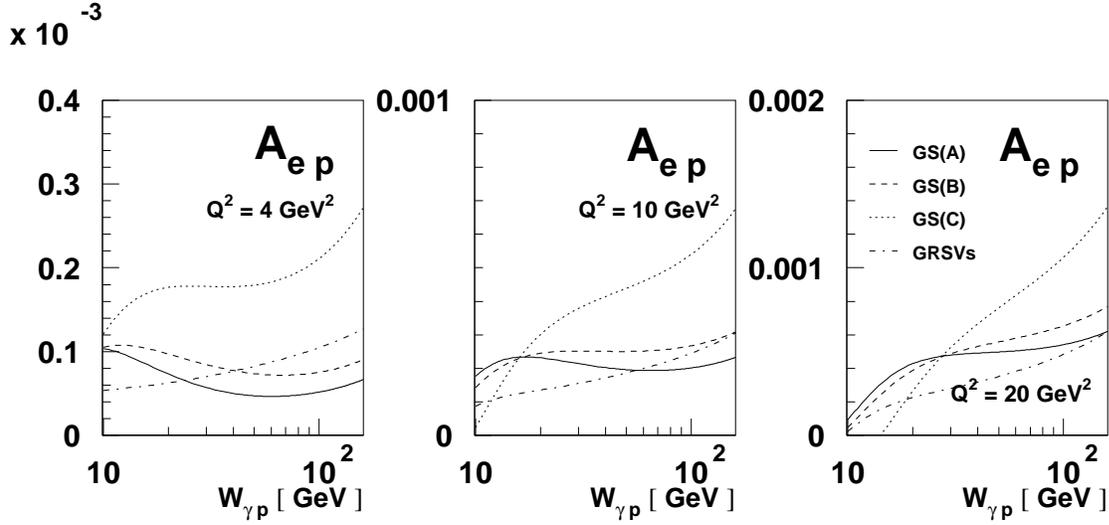,height=15cm,angle=-90}
\caption{
Asymmetries in elastic $\rho$ leptoproduction for different $Q^2$.}
\label{fig4}
\end{center}
\end{figure}

An estimate of the asymmetry in elastic $\rho$ meson electroproduction,
based on the formulae of~\cite{R3} is shown in Fig.~\ref{fig4}. Since
we are only interested in an order-of-magnitude determination of this
asymmetry, two approximations have been made. As suggested in~\cite{R3},
we have used fixed anomalous dimensions for the small $x$ behaviour
of the parton distributions. The integration over the momentum
fractions carried by quark and antiquark is simplified by using the
saddle point method.

All parameterizations of $\Delta G(x,Q^2)$ yield
electron-proton asymmetries of the order of 0.0001 at $Q^2=4$~GeV$^2$, which
rise by about one order of magnitude if  $Q^2$ is increased to 20~GeV$^2$.
It is found that the contribution from the polarized quark sea
(cf.~Fig.~\ref{fig2}) amounts maximally 10\% to the total asymmetry.
Based on the unpolarized measurement of~\cite{h1exp}, we have again estimated
the anticipated errors on this asymmetry, using the same parameters as
above, but $\epsilon_{\rho}=1$. With two bins
in the interval $W_{\gamma p}=40\ldots 140$~GeV, the estimated
error on each point
would be $\pm 0.01$ for $Q^2=10$~GeV$^2$ and $\pm 0.015$ for
 $Q^2=20$~GeV$^2$, which is about one order of magnitude larger than
the expected asymmetry. Thus the $\rho$ meson leptoproduction
asymmetry is clearly not measurable at the HERA collider.

\section{Asymmetries in inelastic heavy vector meson photoproduction
at HERA}
\label{sec:inelth}
\vspace{1mm}
\noindent
In the framework of the color singlet model,
the inelastic photoproduction of a heavy vector meson is described as the
perturbative production of a color singlet $q\bar q$ pair with the quantum
numbers of the quarkonium state \cite{colsing}. The formation of a meson from
the quark pair is a non-perturbative process. Its probability reduces to a
single parameter related to 
the meson wave function at the origin $|\Psi(0)|^2$, which
is thought to be known for $J/\psi$ and $\Upsilon$ families from the leptonic
decay widths $\Gamma(V\to l^+l^-)$. This model yields a satisfactory 
description~\cite{Kraemer} of recent experimental data~\cite{h1exp}, once 
next-to-leading order corrections~\cite{nlocol} to partonic subprocess and 
meson wave function are included. 

An application of the color singlet model to 
vector meson production in polarized photon--hadron collisions 
has been proposed in ref.~\cite{Guillet}. The next-to-leading 
order corrections are presently unknown in the polarized case, one should 
however expect that they partially cancel in the polarization 
asymmetry, the ratio of polarized and unpolarized cross sections. 
In particular, the leading order 
asymmetry is insensitive to the choice of the 
renormalization scale $\mu$ used in the strong coupling constant 
$\alpha_s(\mu)$, while polarized and unpolarized cross sections scale like 
$\alpha_s^2(\mu)$. 

In addition to the color singlet production 
 the so-called color octet mechanism \cite{Braaten} is also widely
discussed in the literature. This model considers the production of a $q\bar q$
pair in a color octet state, i.e.~with ``wrong'' quantum numbers. 
Then, after a
(serial) soft gluon emission this pair is assumed to convert into real meson.
The probabilities for the color octet transitions are not calculable within the
theory and are in fact free model parameters. Their values have been estimated
in ref.~\cite{ChoLeib} from a fit to Tevatron data on $J/\psi$
hadroproduction.
However, the applicability of the color octet model to HERA conditions is
doubtful. Given the transition probabilities (i.e.~the color octet matrix
elements) as estimated in the original paper \cite{ChoLeib}, the color octet
model predictions seem to disagree with the experimental results for $ep$
collisions~\cite{cac}. Several solutions to this disagreement have 
been proposed, the situation is however still inconclusive. Since 
the color singlet model alone yields already a good description~\cite{Kraemer} 
of the 
unpolarized data at HERA, we shall restrict us to this model and 
consistently neglect possible color octet contributions.
\begin{figure}[t!] % fig 6
\begin{center}
~ \epsfig{file=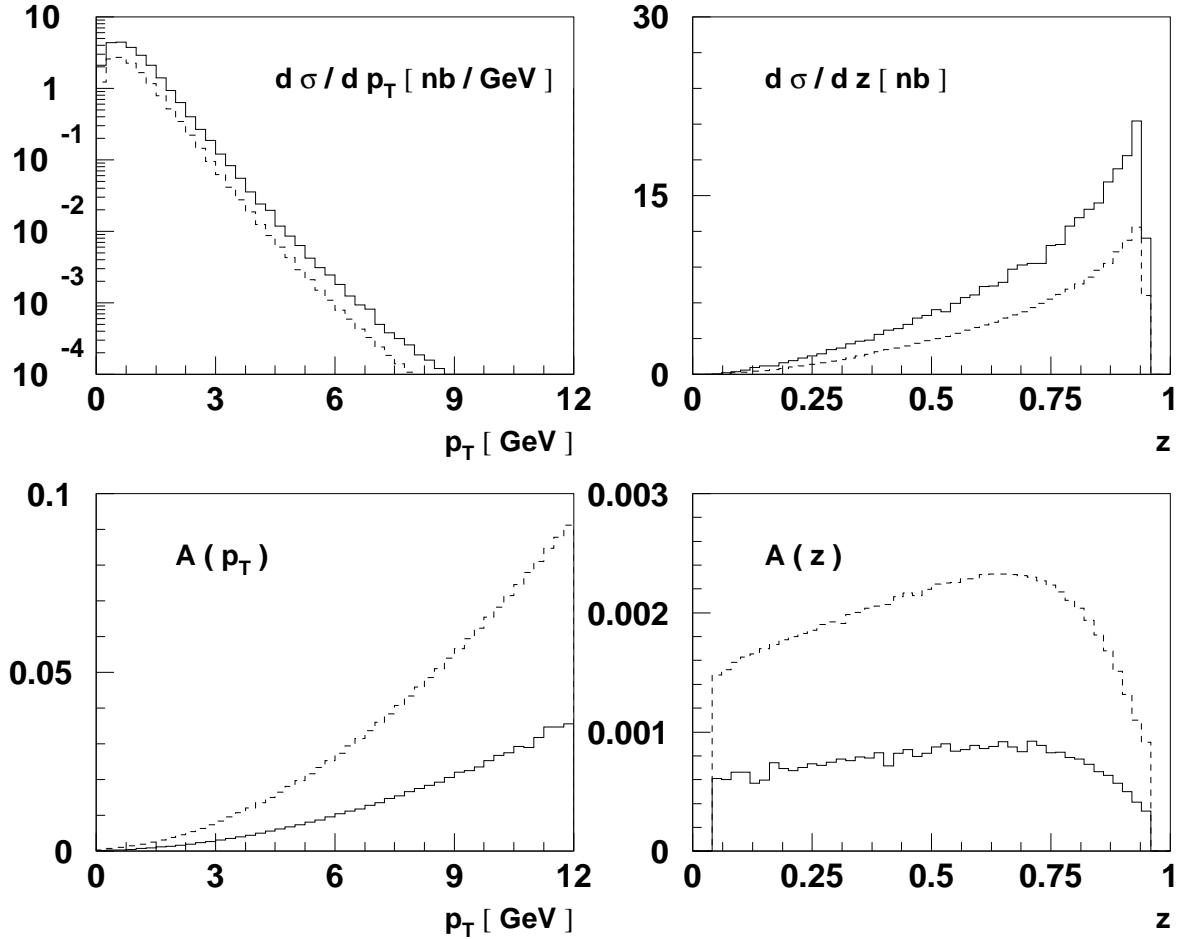,angle=-90,width=16cm}
\caption{ Production cross sections and polarization asymmetries 
in inelastic $J/\psi$ photoproduction at HERA,  $\sqrt s=$ 300~GeV (solid
histogram) and 190~GeV (dashed histogram). Cross sections and 
asymmetries are plotted as function of the $J/\psi$ transverse
momentum $p_T$ (left) and of the photon energy fraction $z$ carried by the 
$J/\psi$ (right). The unpolarized parton distribution functions are taken 
from GRV~\protect{\cite{GRV}}, the polarized ones from GS, set 
(A)~\protect{\cite{GS}}.}
\label{fig6}
\end{center}
\end{figure}

The analyzing power of the partonic subprocess $\gamma+g\to J/\psi+g$
is rather large, but a substantial suppression is 
induced by the ratio  of 
polarized to unpolarized 
gluon densities $\Delta G(x)/ G(x)$ and the photon depolarization factor
$D(y)$, eq.~(\ref{D}).
In the region $p_t<2.5$ GeV (where the cross section is high)
the visible polarization asymmetry for $J/\psi$ photoproduction
at HERA conditions does not exceed 0.002, while its average value stays at
$A\simeq 0.0007$ (see fig.~\ref{fig6}).
 Here we have assumed that no other cuts than the inelasticity
 requirement $z=E_{J/\psi}/E_{\gamma}<0.9$ are applied.
The anticipated statistical errors ($\pm 0.002$ for the same 
parameters as in the previous section but only a 
single bin covering the whole $30$~GeV~$<W_{\gamma p}< 150$~GeV 
range at HERA, cf.~\cite{h1exp}) on a measurement 
of this  asymmetry exceed the value of this asymmetry by 
almost one order of magnitude. 

The kinematical selection of large-$x$, large-$y$ region may give access to
highly polarized gluons and photons, but only at the price of practically
vanishing partonic cross section. For example, the condition $x>0.03$,
$y>0.2$ would lead to $D(y)\,\Delta G(x) / G(x) \gsim 0.04$, but implies 
$\hat\sigma(\gamma+g\to J/\psi+g)$ falling by three orders of magnitude.
The increase in the observable asymmetry is therefore more than 
compensated by the strongly decreasing event statistics. 
Finally, the replacement of the $J/\psi$ meson
by a $\Upsilon$ meson
enhances the asymmetry by approximately one order of magnitude
(i.e.~from $5\cdot 10^{-4}$ to $5\cdot 10^{-3}$) but also yields  
a production rate which is lowered by a
factor of 500 ($\sigma_{\Upsilon}\simeq$ 20 pb versus $\sigma_{J/\psi}\simeq$
10 nb at $\sqrt s = 300$ GeV).
The situation does not significantly improve if HERA is operated 
at $\sqrt s = 180$ GeV. In this context, the measurement of the
polarization asymmetries in inelastic heavy vector meson production
at HERA collider seems infeasible.

\section{Conclusions}
\vspace{1mm}
\noindent
In this report we have investigated the asymmetries in the
production of vector mesons in polarized electron-proton collisions at HERA.
Our estimates are based on the extension of perturbative models for
unpolarized
elastic and inelastic production of vector mesons to the polarized
case. Numerical studies of
elastic and inelastic $J/\psi$ photoproduction and
elastic $\rho$ electroproduction have shown that the prospects
for a measurement of asymmetries in vector meson production are rather poor,
since the magnitude of the expected asymmetries is usually similar or smaller
than the expected statistical errors on the measurement. The lowering of the
HERA proton beam energy would lead to an enhancement of all vector meson 
production asymmetries, this enhancement is however not sufficient to 
make these asymmetries experimentally accessible. \vspace{0.4cm}

\noindent
{\bf Acknowledgement:} The work of MGR was supported by the INTAS grant 
93-0079-EXT and by the Volkswagen-Stiftung.

\end{document}